\documentclass[conference]{IEEEtran}

\usepackage{amsmath}
\usepackage{cite}
\usepackage{graphicx}
\usepackage{amsfonts}
\usepackage{latexsym}
\usepackage{subfigure}
\usepackage{algorithm}
\usepackage{algorithmic}
\usepackage{color}
\usepackage{times}
\usepackage{epsfig, graphics}
\usepackage{bm}
\usepackage{subfigure}
\usepackage{graphicx,wrapfig,lipsum}
\begin{document}

\newtheorem{theorem}{Theorem}
\newtheorem{lemma}{Lemma}
\newtheorem{conjecture}{Conjecture}
\newtheorem{corollary}{Corollary}
\newtheorem{definition}{Definition}
\newtheorem{scheme}{Scheme}
\newcommand{\argmax}{\arg\!\max}
\newcommand{\rev}[1]{{\color{black}#1}} 
\newcommand{\pound}{\operatornamewithlimits{\gtrless}}
\IEEEoverridecommandlockouts

\title{IoT Network Security from the Perspective of Adversarial Deep Learning}



\author{\IEEEauthorblockN{Yalin E. Sagduyu$^*$, Yi Shi$^{*\dagger}$, and Tugba Erpek$^{*\ddagger}$}
\IEEEauthorblockA{$^*$Intelligent Automation, Inc., Rockville, MD 20855, USA \\ $^{\dagger}$Department of Electrical and Computer Engineering, Virginia Tech, Blacksburg, VA 24061, USA \\ $^{\ddagger}$Department of Electrical and Computer Engineering, Hume Center, Virginia Tech, Arlington, VA 22203, USA \\
Email: \{ysagduyu, yshi, terpek\}@i-a-i.com
}

\thanks{This effort is supported by the U.S. Army Research Office under contract
W911NF-17-C-0090. The content of the information does not necessarily
reflect the position or the policy of the U.S. Government, and no official endorsement should be inferred.}
}


\maketitle

\begin{abstract}
Machine learning finds rich applications in Internet of Things (IoT) networks such as information retrieval, traffic management, spectrum sensing, and signal authentication. While there is a surge of interest to understand the security issues of machine learning, their implications have not been understood yet for wireless applications such as those in IoT systems that are susceptible to various attacks due the open and broadcast nature of wireless communications. To support IoT systems with heterogeneous devices of different priorities, we present new techniques built upon adversarial machine learning and apply them to three types of over-the-air (OTA) wireless attacks, namely jamming, spectrum poisoning, and priority violation attacks. By observing the spectrum, the adversary starts with an exploratory attack to infer the channel access algorithm of an IoT transmitter by building a deep neural network classifier that predicts the transmission outcomes. Based on these prediction results, the wireless attack continues to either jam data transmissions or manipulate sensing results over the air (by transmitting during the sensing phase) to fool the transmitter into making wrong transmit decisions in the test phase (corresponding to an evasion attack). When the IoT transmitter collects sensing results as training data to retrain its channel access algorithm, the adversary launches a causative attack to manipulate the input data to the transmitter over the air. We show that these attacks with different levels of energy consumption and stealthiness lead to significant loss in throughput and success ratio in wireless communications for IoT systems. Then we introduce a defense mechanism that systematically increases the uncertainty of the adversary at the inference stage and improves the performance. Results provide new insights on how to attack and defend IoT networks using deep learning.
\end{abstract}


\section{Introduction}

\emph{Machine learning} has been successfully applied in diverse areas such as image processing, natural language processing (NLP), and autonomous driving. 
Emerging hardware resources have increased the computing power and thus the amount of training data that can be processed by machine learning has grown significantly. 
In particular, \emph{deep learning} that corresponds to training sophisticated structures of deep neural networks has achieved significant progress in improving the decision making for various detection, classification, and prediction tasks. Consequently, machine learning can make better decisions such as recognizing human face, understanding natural language (including social media that may not exactly follow grammar rules), recognizing traffic signals for self-driving cars, and beating professional gamers in traditional games (e.g., Google's AlphaGo \cite{Silver}) and computer games \cite{openai}.

Internet of Things (IoT) systems rely on various detection, classification, and prediction tasks \cite{Mohammadi18, Li18} to learn from and adapt to the underlying spectrum environment characterized by heterogeneous devices, different priorities, channel, interference, and traffic effects. Machine learning has emerged as a powerful tool to perform these tasks in an automated way by learning from the complex patterns of the underlying wireless communication dynamics. Examples of machine learning applied to wireless communications include modulation recognition  \cite{OShea2016}, spectrum sensing \cite{Thilina2013, Kemal2018, Lee2017}, interference management \cite{Erpek2018ICC}, and routing \cite{Hu2010}. IoT systems can benefit from cognitive radio capabilities (potentially empowered by machine learning) for efficient use of limited spectrum resources available to IoT applications \cite{Khan17}.

Recently, there has been a surge of efforts to apply machine learning to \emph{wireless security}, including spoofing attacks \cite{spoofing}, jamming attacks  on data transmission \cite{Yi2018,Tugba2018}, and other attacks that target spectrum sensing \cite{Yi2018Milcom} and signal classification \cite{Sadeghi2018} tasks. In particular, IoT system security benefits from machine learning to identify devices \cite{Miettinen17, Meidan2017}, authenticate signals \cite{Saad2018}, and detect anomalies \cite{Canedo2016}. With growing applications of machine learning, it is necessary to understand the underlying security threats that target machine learning itself. \emph{Adversarial machine learning} has emerged as a field to systematically analyze the security implications of machine learning in the presence of adversaries \cite{Barreno2006, Huang2011, Murat2018}. Traditionally applied to data domains other than wireless communications, various attacks studied by adversarial machine learning include the following.
\begin{itemize}
\item \emph{Exploratory (or inference) attack} where the adversary aims to understand how the target machine learning algorithm, e.g., a classifier, works \cite{Fredrikson,Shi17:HST,Yi2018}.
\item \emph{Evasion attack} where the adversary aims to fool a machine learning algorithm into making wrong decisions  \cite{Biggio,Yi2018}.
\item \emph{Causative (or poisoning) attack} where the adversary aims to provide incorrect training data for a machine learning algorithm to (re)train itself \cite{Pi}.
\end{itemize}
These attacks can be launched separately or combined, e.g., causative and evasion attacks can be launched building upon the inference results of an exploratory attack \cite{Shi18:book}. These attacks have been broadly applied against image and text classifiers \cite{ShiMilcom, HST2018, ISSPIT2018}. 

We present the application of adversarial machine learning in IoT systems by mapping the techniques developed for exploratory, evasion and causative attacks to different wireless attacks, namely \emph{jamming}, \emph{spectrum poisoning} and \emph{priority violation attacks} in IoT networks. Adversarial deep learning was applied in the test phase to jamming the data transmission phase in \cite{Yi2018, Tugba2018} and jamming the spectrum sensing phase in \cite{Yi2018Milcom}. In this paper, we use these attacks as benchmarks and extend them to the (re)training phase to improve energy efficiency and stealthiness of these attacks. We add the new attack of priority violation built upon adversarial deep learning. 

The main difference of these attacks from the conventional wireless attacks (such as jamming, e.g., \cite{Berry2009, Sagduyuwireless2009, Berry2011}) is that the adversary directly targets the machine learning algorithm (such as the one for spectrum sensing) used in IoT communications. As the first step, the adversary senses the spectrum and infers the transmission outcomes of an IoT device by monitoring the presence of transmission acknowledgements (ACKs) over the channel. One important difference between the traditional use of adversarial machine learning (such as querying an application programming interface (API) for computer vision) and its wireless application is that a wireless adversary can only partially observe the features (sensing result) and labels (channel status) to train a classifier. Using these features and labels (that may differ from those of the IoT transmitter due to different channels), the adversary trains a classifier using a \emph{deep neural network}, as a form of an \emph{exploratory attack} to predict whether there will be a successful transmission at any given time. The adversary uses this classifier to launch the following attacks next.
\begin{itemize}
	\item \emph{Jamming}: If the adversary predicts that there will be a successful transmission, it jams the data transmission such that the receiver cannot decode the transmitted data.
	\item \emph{Spectrum poisoning}: The adversary jams the sensing phase with the goal of changing the features that describe the current channel status and forcing the transmitter into making wrong transmit decisions. Therefore, it can be treated as an \emph{evasion attack}. Note that this attack differs from the spectrum sensing data falsification (SSDF) attack \cite{Wang2015}, since the adversary does not participate in cooperative spectrum sensing and does not try to change the estimated channel labels directly as in the SSDF attack. Instead, \emph{the adversary injects adversarial perturbations over the air to the channel}.
	\item \emph{Violating priorities}: The IoT transmitter with low priority aims to avoid interference to IoT users with higher priority. The adversary transmits during the sensing phase by pretending to have higher priority. This attack forces the transmitter into making wrong 
	decisions and can be treated as an \emph{evasion attack}. 
	It is similar to the primary user emulation attack \cite{Nguyen2015} that has not been studied yet from the adversarial machine learning point of view. 
\end{itemize}

Another difference of wireless attacks in IoT systems from the conventional use of adversarial machine learning is that a wireless adversary cannot directly manipulate the test or training data input to a classifier but can only indirectly try to do this by jamming the channel during the data transmission or sensing phases.  In all three wireless attack types, 
an IoT transmitter may also \emph{retrain} and improve its classifier with additional training data collected. 
Then the adversary can further launch an attack to change the
transmission outcomes (labels) or the channel status (features) to be used as part of the input to the retraining process and thus reduce the transmitter's performance, which corresponds to a \emph{causative attack}.

For all these wireless attacks, we demonstrate the success of the unique techniques from adversarial machine learning to infer the target classifier used for spectrum sensing and transmit decisions, and manipulate its input in terms of test and training data. Results clearly demonstrate the major performance loss of IoT systems in terms of classification error, throughput and success ratio for data offloading, and raise the need to develop defense mechanisms against adversarial machine learning.
Thus, we design a \emph{defense} approach, where the IoT transmitter selectively makes some wrong actions such that the exploratory attack cannot succeed. 
Consequently, subsequent attacks cannot be successful either.
A tradeoff for this defense approach is on the number of defense actions.
If this number is small, there is only a limited impact on attacks while if this number is large, the transmitter's performance drops even when there is no attack.
We describe this problem as a Stackelberg game and consider the defense for priority violation attack as one example.
Results show that with the optimal number of defense actions, the throughput of the IoT transmitter can be increased significantly.


The rest of the paper is organized as follows.
Section~\ref{sec:related} discusses related work. 
Section~\ref{sec:model} presents the deep learning model and overviews adversarial machine learning.
Section~\ref{sec:app_wireless} describes the IoT system setting to apply adversarial machine learning.
Section~\ref{sec:jam}, \ref{sec:sense}, and \ref{sec:pu} discuss the jamming, spectrum poisoning, and priority violation attacks, respectively. Section~\ref{sec:discuss} presents a defense approach against adversarial machine learning.
Section~\ref{sec:conclusion} concludes the paper.

\section{Related Work}
\label{sec:related}

Security issues for IoT systems have been widely studied in the literature with machine learning applications, e.g., \cite{Miettinen17, Meidan2017, Saad2018, Canedo2016}.
In \cite{Meidan2017}, a multi-stage meta classifier was trained by machine learning to first distinguish between traffic generated by IoT and non-IoT devices, and then determine IoT device class.
In \cite{Miettinen17}, machine learning was used to automatically identify the types of devices being connected to an IoT network and enable enforcement of rules for constraining the communications of vulnerable devices to minimize damage resulting from their compromise. 
In \cite{Saad2018}, deep learning was used to detect data injection and eavesdropping in IoT devices. 
In \cite{Canedo2016}, machine learning was applied within an IoT gateway to detect anomalies in the data sent from the edge devices.



\begin{table*}
	\caption{Adversarial machine learning attacks on wireless communications.}
	\centering
	{\small
		\begin{tabular}{c|c|c|c|c|c|c}
			Attack & \multicolumn{3}{c|}{Attack in test phase} & \multicolumn{3}{c}{Attack in (re)training phase} \\ \cline{2-7}
			& Jamming & Spectrum poisoning & Priority violation & Jamming & Spectrum poisoning & Priority violation \\ \hline
			\cite{Yi2018, Tugba2018} & \checkmark & --- & --- & --- & --- & ---\\ \hline
			\cite{Yi2018Milcom} & --- & \checkmark & --- & --- & --- & --- \\ \hline
			This paper & \checkmark & \checkmark & \checkmark & \checkmark & \checkmark & \checkmark
		\end{tabular}
	}
	\label{table:related}
	\vspace{-5mm}
\end{table*}

Adversarial machine leaning has started finding applications in wireless communications. In \cite{Sadeghi2018}, white-box and black-box attacks were designed by changing the input data to a modulation classifier based on deep learning. Our approach is different as it targets data transmission and spectrum sensing phases, and explicitly describes how to manipulate the input data in test and training phases. Jamming attack and defense mechanisms were developed in \cite{Yi2018, Tugba2018} with deep learning, while \cite{Yi2018Milcom} studied a spectrum poisoning attack. In these studies, retraining process was not considered. In this paper, we provide a common adversarial machine learning framework including both the jamming and spectrum poisoning attacks, along with priority violation attacks, and extended the attack space from test phase to (re)training phase for IoT systems. 
Table~\ref{table:related} summarizes applications of adversarial machine learning to wireless security, where a `\checkmark' mark indicates that the corresponding problem is considered in a paper.

\section{Deep Neural Network Model and Overview of Adversarial Machine Learning}
\label{sec:model}
\subsection{Deep Neural Network Model}

Machine learning makes decisions by learning from data, without being explicitly programmed on how to make decisions. Applications of machine learning can be categorized as classification (if decisions are discrete values) or regression (if decisions are continuous values). We focus on classifiers in this article. A classifier can be built by using the training data that has been labeled, where each sample in the training data is described by a number of features and is associated with a label (or class). Once trained, a classifier can determine a label for any given unlabeled sample. The performance of a classifier can be measured by the error of the determined labels for some test data. Then an adversary can target the classifier in terms of learning how the underlying algorithm functions or manipulating training or test data to reduce its performance.

For a classification problem, each sample $s_i$ is modeled by a number of features $(f_{i1}, f_{i2}, \cdots, f_{iM})$.
Denote $L(s_i)$ as the actual label for sample $s_i$ and $C(s_i)$ as the label determined by a classifier $C$.
Ideally, a perfect classifier should always produce $C(s_i) = L(s_i)$.
In reality, a classifier may not always make the correct decision, i.e., it may produce $C(s_i) \ne L(s_i)$ for some sample $s_i$.
We can use a set of test data to measure the performance of a classifier.
Suppose that the size of test data is $n_{test}$ and the number of samples with $C(s_i) \ne L(s_i)$ is $n_{error}$.
Then the error probability is calculated as $\frac{n_{error}}{n_{test}}$.


There are various machine learning algorithms such as Naive Bayes, support vector machine, 
neural networks and deep learning, to train a classifier. We apply deep learning at both the transmitter and the adversary to capture the complex interactions of channel, traffic, and actions of the IoT transmitter and the adversary.
Deep learning extends neural networks with more hidden layers in addition to the input and output layers. In this paper, we consider a  feedforward neural network (FNN). 
Each layer consists of neurons.
The neurons in the input layer accept features of a sample $s_i$ as inputs.
Each neuron processes its input data by an activation function and provides its output to neurons in the next layer. Neurons in the output layer provides various outputs, including label $C(s_i)$ for sample $s_i$ and a score that measures the confidence of classification.
The mathematical function at each neuron is not predetermined and consists of weights and biases. 
A number of training data samples is used to train the deep neural network, i.e., optimally determine its weights and biases such that the output labels are close to the real labels. Details of the deep neural networks are discussed in Section~\ref{sec:app_wireless}.

\subsection{Overview of Adversarial Machine Learning}

\begin{figure}
	\centering
	\includegraphics[width=0.75\columnwidth]{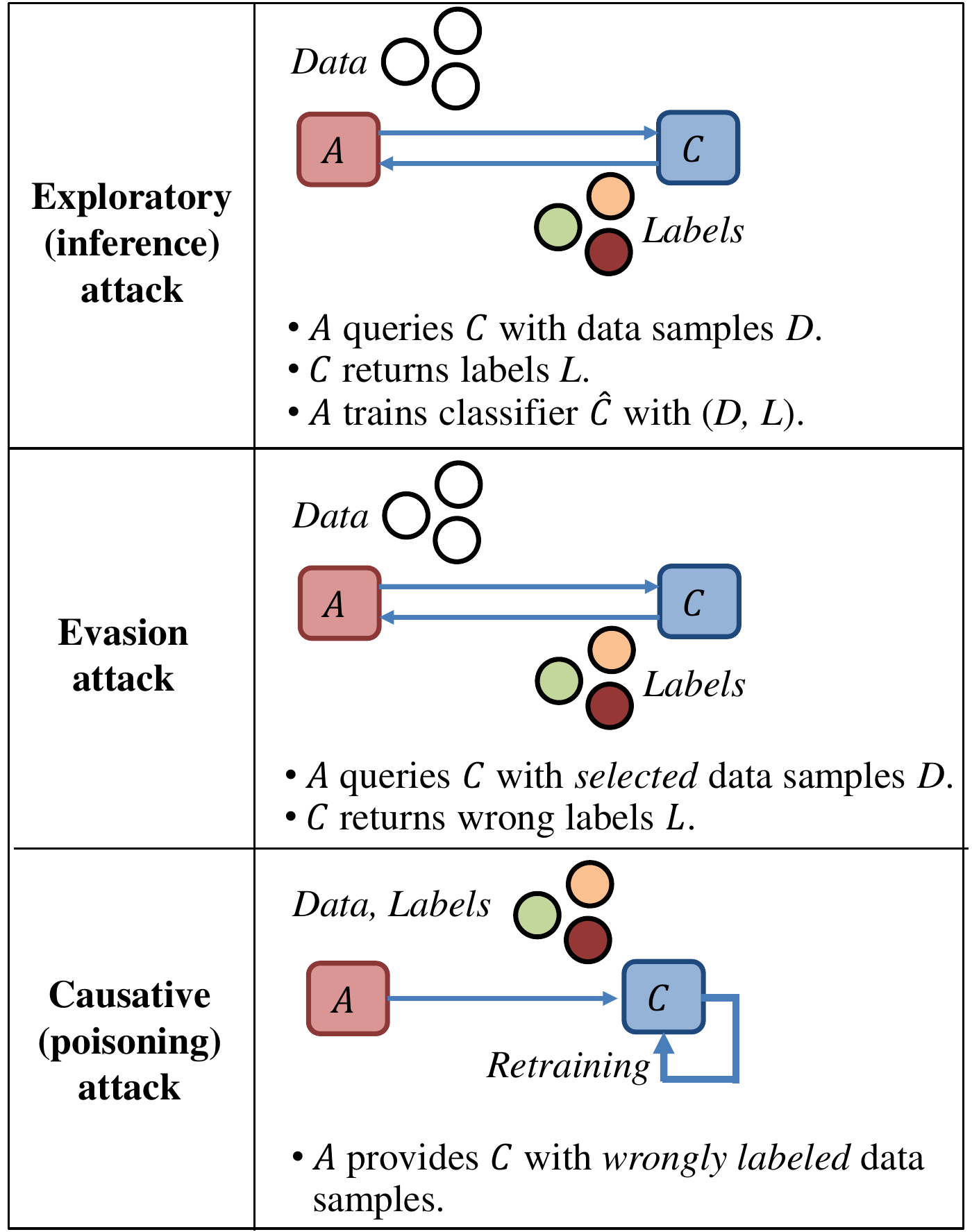}
	\caption{Types of attacks under adversarial machine learning.}
\label{fig:aml}
	\vspace{-5mm}
\end{figure}

Adversarial machine learning has been studied for different data domains mostly focused on image and text processing \cite{Murat2018}. 
In this paper, we consider an IoT transmitter $T$ using its pre-trained classifier $C$ to detect idle channels for data transmissions.
The adversary $A$ applies the following three types of attacks (see Figure~\ref{fig:aml}) on the target classifier $C$. 
\begin{itemize}
\item \emph{Exploratory attack}. The adversary aims to build a classifier $\hat C$ similar to the target classifier $C$.
That is, for most samples, we have $\hat C(s_i) = C(s_i)$.
Such an attack can be a white-box attack, where the adversary knows the classifier algorithm and/or training data, or a black-box attack, where the adversary does not have any knowledge on the classifier algorithm or training data.
We will focus on a black-box attack in this paper.

\item \emph{Evasion attack}. The adversary aims to determine the data samples that are most likely to be misclassified by the target classifier. One example of such attack is that an adversary wants to generate spam emails that cannot be identified as spam.

\item \emph{Causative attack}. If the original classifier is trained with limited data, it may be retrained with additional training data. However, the adversary can attack this retraining process by providing incorrect training data.
\end{itemize}
These attacks can be applied separately or jointly. We will describe their application of these attacks in wireless communications for an IoT network in Section~\ref{sec:app_wireless}.

\section{Adversarial Deep Learning for Wireless Attacks}
\label{sec:app_wireless}
There are important differences between the traditional use of adversarial machine learning (such as querying a computer vision API) and its application in wireless communications.
\begin{itemize}
	\item In the wireless setting, the adversary cannot query a transmitter's classifier directly with an input sample and receive its label output. Instead, the adversary can only observe the transmitter's behavior over a channel, i.e., the adversary can only obtain noisy samples of input data that may be different from what the transmitter observes due to different channel and interference effects. Moreover, the adversary predicts the outcome of transmissions (`ACK' or `no ACK') while the transmitter predicts channel status (`idle' or `busy'). Hence, the labels of their classifiers will also differ.
	
	\item In the wireless setting, the adversary cannot directly manipulate the test or training data input to a classifier (such as selecting the exact input data to the classifier or flipping training data labels in the training process). Instead, the adversary can only decide on how to make transmissions that manipulate the received signals so that the test or training process is indirectly fooled.
\end{itemize}

To provide insights into the application of adversarial machine learning to wireless communications, we consider the following setting.
A machine learning-based classifier (namely, a deep neural network) is used at an IoT transmitter $T$ to analyze the spectrum sensing data and identify idle channels for data transmission. Packets arrive at a background transmitter $B$ randomly according to the Bernoulli process with rate $\lambda = 0.8$ (packet/slot). Whenever $B$ is idle and it has packet(s) to transmit, it is activated with certain probability to continue transmitting until it empties its queue. Therefore, it is possible that a continuous period of slots remains busy depending on the number of previous idle slots. As a result, there is a temporal correlation among channel busy/idle states and therefore it is necessary for both $T$ and $A$ to observe the past channel status over a time period to predict the current channel status.
\rev{To simplify the discussion, we assume a single background transmitter (potentially, another IoT transmitter with higher priority).
For the general case of multiple background transmitters, channel busy/idle status is determined by the aggregated signals from all these transmitters.
This setting does not affect the attack/defense approaches developed in the paper.}

$T$ uses a time slot structure with three phases, the \emph{sensing} phase, the \emph{transmission} phase, and the \emph{feedback} phase. $T$ collects channel status  during the sensing phase and then applies a machine learning algorithm to predict whether the channel is busy or idle. In this paper, we assume that $T$ applies a deep neural network.
If the channel is predicted as idle, $T$ transmits its data during the transmission phase.
The receiver $R$ sends ACK in the feedback phase if data is received. $T$ optimizes the hyperparameters of its deep neural network (e.g., number of layers) by solving the optimization problem {\it HyperT} that minimizes
$ \max\{e_{_{MD}}(C), e_{_{FA}}(C)\}$, where 
$e_{_{MD}}(C)$ is the misdetection probability of classifier $C$ that a busy slot is classified as idle and and $e_{_{FA}}(C)$ is the false alarm probability of $C$ that an idle slot is classified as busy. 


We end up with the following structure of deep neural network implemented with TensorFlow. FNN is trained with the backpropagation algorithm by using the cross-entropy as the loss function. Number of hidden layers is 3. Number of neurons per hidden layer is 100. Rectified linear unit (ReLU) is used as activation function at hidden layers. ReLU performs the $f(x) = \max(0,x)$ operation on input $x$. Softmax is used as the activation function at the output layer. Softmax determines the $i$th entry of $f(\bm{x})$ by performing $f(\bm{x})_i = {e^{x_i}}/{\sum_j e^{x_j}}$ on input $\bm{x}$. Batch size is 100. Number of training steps is 1000.


The adversary $A$ performs a black-box \emph{exploratory attack} to predict the outcome (ACK or not) of transmissions. \rev{Without knowing the classifier of $T$, $A$ senses the channel status, monitors ACKs (labels) over certain period of time and using the collected data to train another classifier. }
Note that the adversary needs to detect the presence of an ACK but does not need to decode its message. ACK messages have certain features that make it easier for the adversary to distinguish them from data messages. First, ACK messages are shorter. Second, they follow a data transmission. Third, the  signal-to-noise-ratio (SNR) to decode them is much smaller than that for data messages. Consequently, $A$ can reliably detect if there is an ACK message transmission that is distinct from a data transmission. Furthermore, by observing inter-arrival times of ACKs, $A$ can determine $T$'s time slot structure including start and end point, duration and its decomposition to sensing, transmission, and feedback periods.



The optimization problem at adversary $A$ is similar to Problem {\it HyperT} and thus is omitted. 

\begin{figure}
	\centering
	\includegraphics[width=0.75\columnwidth]{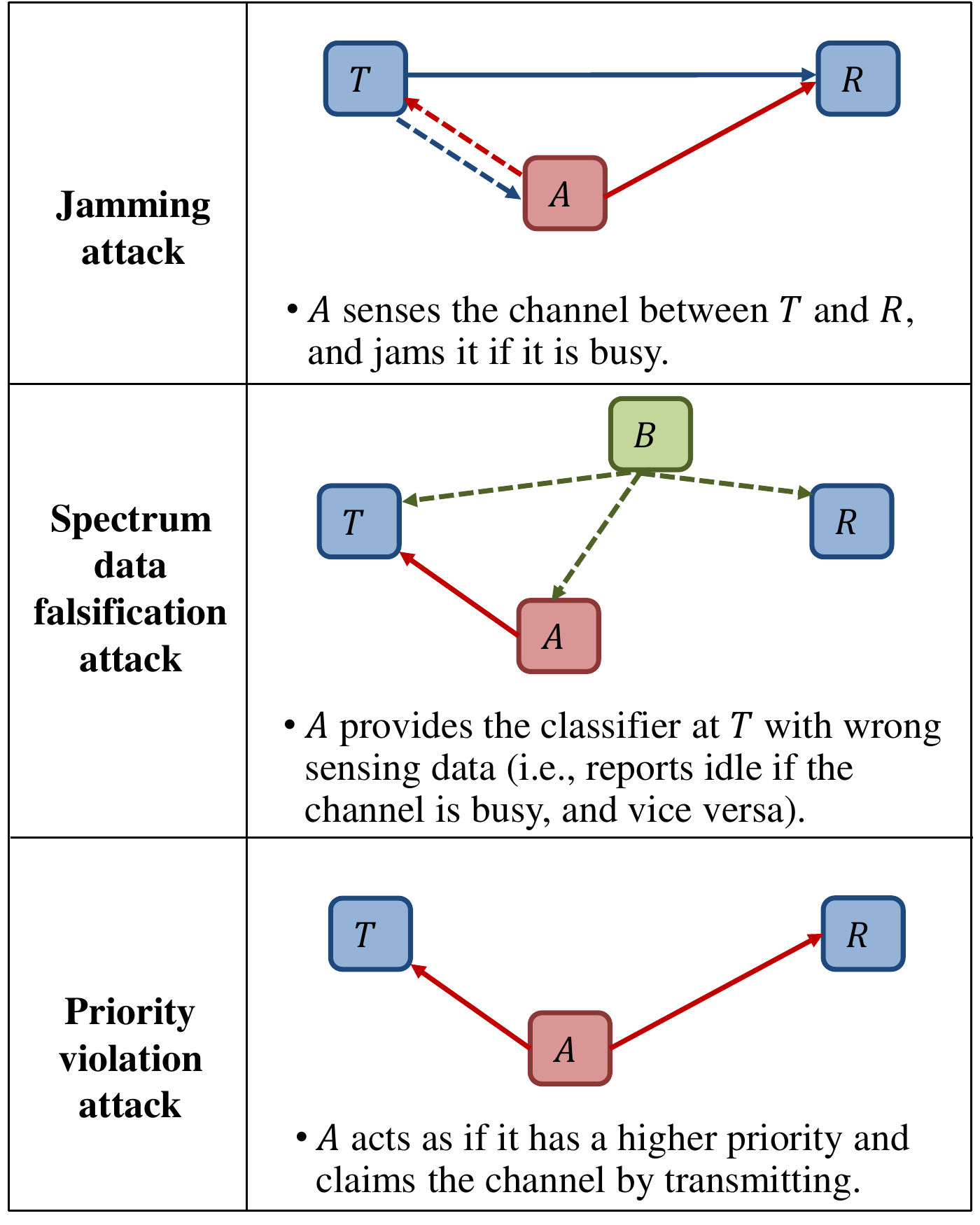}
	\caption{Types of attacks on wireless communications. }
	\label{fig:aml2}
	\vspace{-5mm}
\end{figure}

After the exploratory attack, $A$ analyzes the channel access behavior of $T$ and launches further attacks.
We consider three types of attacks on wireless communications, as illustrated in Figure~\ref{fig:aml2}, by using results from the exploratory attack.

\begin{table*}
	\caption{Jamming attack based on adversarial machine learning.}
	\centering
	{\small
		\begin{tabular}{c|c|c|c|c}
			Attack & False alarm & Misdetection & Normalized throughput & Success ratio \\ \hline
			No attack & 2.68\% & 1.03\% & 96.43\% & 95.58\% \\ \hline
			Jamming attack & 22.69\% & 70.25\% & 22.70\% & 23.12\% \\ \hline
			Jamming attack on retraining & 53.10\% & 31.96\% & 46.40\% & 85.00\% 
		\end{tabular}
	}
	\label{table:jamming}
	\vspace{-5mm}
\end{table*}

\begin{itemize}
\item \emph{Jamming attack}: In a typical jamming attack, the adversary transmits if it predicts a transmission and if there is indeed a transmission, the adversary jams this transmission. The jamming attack in this paper is different. $A$ does not predict whether the channel is idle (and there will be a transmission attempt) and jam all transmissions. Instead, it predicts and jams successful transmissions only, which is more energy efficient than typical jamming attacks. 
Moreover, $T$ may retrain its classifier based on the feedback from $R$.
In this case, $A$ can jam transmissions to change the input training data such that the updated classifier actually becomes worse, which corresponds to a \emph{causative attack}.

\item \emph{Spectrum Poisoning Attack}: $A$ jams the sensing phase.
By doing so, the features to describe the current channel status are changed and thus the decision made by machine learning may also be changed, which corresponds to an \emph{evasion attack}.
$A$ transmits during the sensing phase to inject adversarial perturbations to the channel. The goal of $A$ is to fool $T$'s classifier into making wrong transmit decisions. 
That is, $T$ may transmit in a busy channel or may not transmit in an idle channel, which reduces the communication performance.
Again, $T$ may retrain its classifier and $A$ can manipulate features in the retraining process.
By using the manipulated features to retrain a classifier, $T$'s performance may actually drop, which corresponds to a \emph{causative attack}.

\item \rev{\emph{Priority Violation Attack}:  There is the background traffic from a user with higher priority that follows a certain pattern.
Once a transmission from a high-priority user is detected, the transmitter with lower priority backs off for several time slots to avoid interference to such transmissions. $A$ again jams the sensing phase, but now its objective is to pretend to be a user with high priority.} Due to the backoff mechanism, $A$ can reduce the number of attacks and thus consume less energy than the spectrum poisoning attack.
When $T$ retrains its classifier, $A$ increases (false) high-priority user activities in the input to the retraining process as a form of \emph{causative attack}.
\end{itemize}

\begin{figure}
	\centering
	\includegraphics[width=0.95\columnwidth]{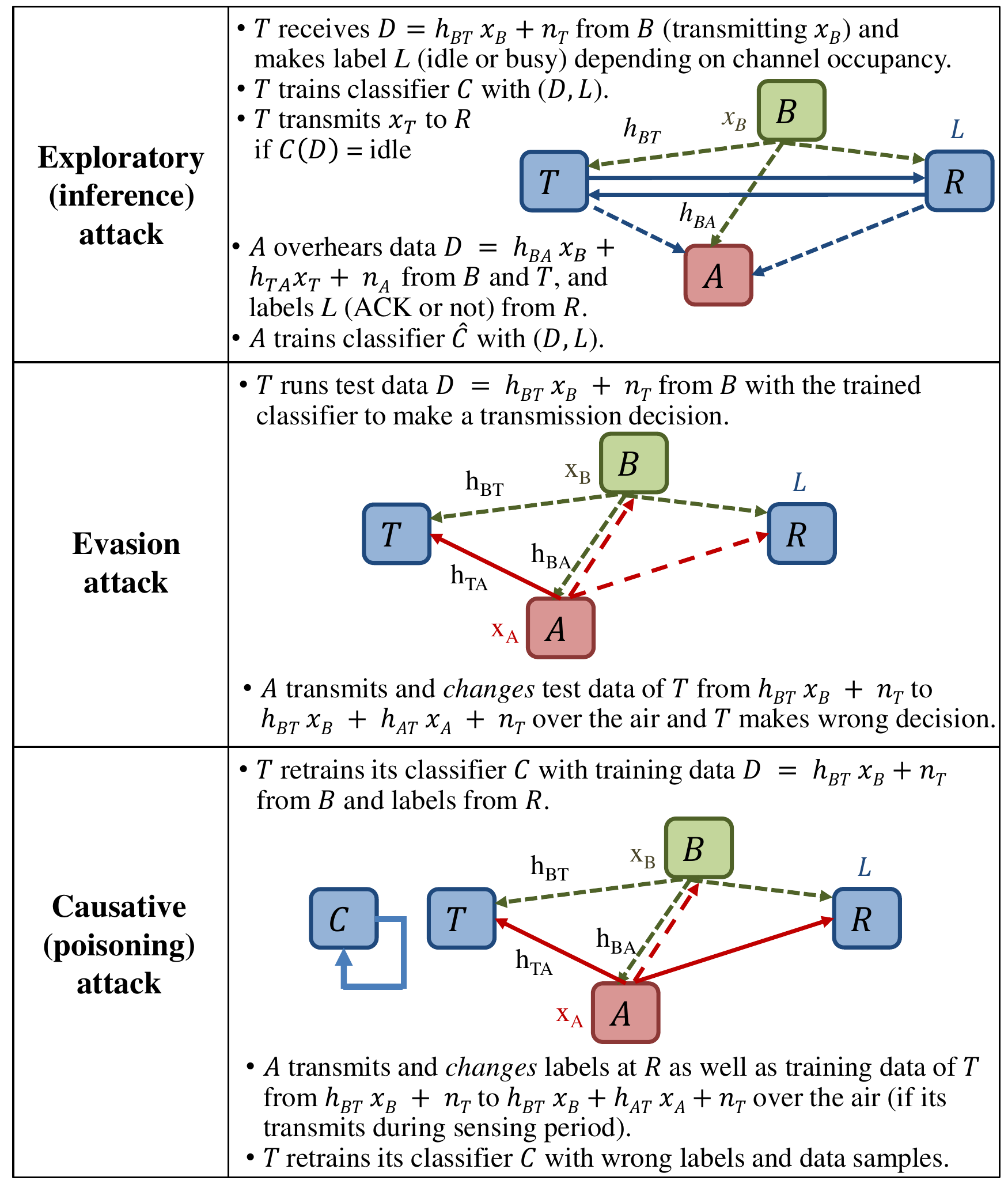}
	\caption{Adversarial machine learning in wireless communications.}
	\label{fig:AML_wireless}
	\vspace{-5mm}
\end{figure}

Figure~\ref{fig:AML_wireless} summarizes the unique properties when adversarial machine learning is applied in wireless communications for an IoT network.
\rev{Note that for causative attack, $A$ can identify $T$'s re-training phase when re-training is launched periodically.
$A$ first determines the end time of re-training phase by observing the accuracy changes due to updated $T$.
Then $A$ further determines the length of re-training phase by adjusting the length of its causative attack periods.}
In the following sections, we will visit each type of wireless attack and show how to apply adversarial machine learning.

\section{Jamming Attack}
\label{sec:jam}
Adversarial deep learning to jam the test phase of data transmission was considered in \cite{Yi2018, Tugba2018}. We first use this attack as a benchmark and then extend this attack to the (re)training phase to improve energy efficiency and stealthiness of the attack. 

We consider the scenario that an IoT transmitter $T$ senses the channel status (`idle' or `busy') and transmits when channel is sensed as idle.
The sensing result is either noise plus interference if channel is busy or noise if channel is idle.
We assume that noise $n_{_T}$ at $T$ follows the Gaussian distribution.
Interference is from some transmitter $B$ (unknown to $T$).
Denote transmit power as $P_B$ and channel gain from $B$ to $T$ as $h_{_{BT}}$.
Then we have the mean interference power as $h_{_{BT}} P_B$.
We assume that interference follows the Gaussian distribution. $T$ applies deep learning to build a classifier $C$ that uses the most recent $n_{new}$ channel sensing results to determine the channel status (`idle' or `busy'). If channel is idle in a time slot, $T$ transmits data with power $P_T$ and receives feedback ACK (if data is received by receiver $R$). $R$ can receive data if the SINR at $R$ is not less than a threshold $\beta$.


\begin{table*}
	\caption{Spectrum poisoning attack based on adversarial machine learning.}
	\centering
	{\small
		\begin{tabular}{c|c|c|c|c}
			Attack & False alarm & Misdetection & Normalized throughput & Success ratio \\ \hline
			No attack & 2.68\% & 1.03\% & 96.43\% & 95.58\% \\ \hline
			Poisoning attack & 91.96\% & 1.29\% & 8.04\% & 64.29\% \\ \hline
			Poisoning attack on retraining & 65.18\% & 29.64\% & 46.43\% & 33.77\% 
		\end{tabular}
	}
	\label{table:poisoning}
\end{table*}

There is an adversary $A$ that launches a jamming attack in three steps. In the first step, $A$ senses channel (noise plus interference or noise) and the transmission outcome (ACK or not) over a period of time.
The collected data is used as training data to build a classifier $\hat C$ in the second step (by using $n_{new}$ most recent sensing results aggregated for each data sample).
This corresponds to an \emph{exploratory attack}.
However, there are some subtle differences from the conventional use of exploratory attack discussed in Section~\ref{sec:model}:
\begin{itemize}
\item The features are different. Note that the power levels sensed at $T$ and $A$  are usually different due to random channel effects. Thus the features (most recent $n_{new}$ sensing results) are different at $T$ and $A$.

\item The labels are different. $T$ aims to predict whether a channel is idle or busy.
$A$ cannot directly query the classifier and obtain labels. Instead, $A$ obtains labels in terms of overheard feedbacks and aims to predict whether there is a successful transmission.
\end{itemize}
Due to these differences, the classifier $\hat C$ may not be similar to $C$.
Instead, $\hat C$ will predict the outcome of actions based on $C$ using the sensing data at $A$.
In the third stage, $A$ starts jamming with power $P_A$ if it predicts that $T$ will transmit and this transmission will be successful.
With the additional interference from $A$, a transmission may fail (once the SINR at $R$ is less than threshold $\beta$).


The locations of $B$, $T$,  $R$, and $A$ are $(0,15)$, $(-10,0)$, $(0,0)$, and $(10,0)$, respectively. 
Channel gain $h$ is modeled as $h = d^{-2}$, where $d$ is the distance between two nodes.
We assume that the mean noise power is normalized as $1$.
The (normalized) powers at $B$, $T$,  $R$, and $A$ are $P_B=P_T=P_A=1000$.
The SINR threshold is $\beta=3$.
The classifier $C$ is built on $1000$ training data samples (see optimization problem {\it HyperT} in Section~\ref{sec:model}) and applied on $500$ test data samples. $n_{new} = 10$ most recent sensing results are aggregated as one input sample. 
The false alarm (idle channel is determined as busy) and misdetection (busy channel is determined as idle) probabilities are measured as $e_{_{FA}}(C)=2.68\%$ and $e_{_{MD}}(C)=1.03\%$, respectively.
The achieved throughput is normalized by the ideal throughput (every idle channel is detected for transmission).
The normalized throughput is measured as $96.43\%$. In addition, the success probability of attempted transmissions is measured as $95.58\%$. 
These results show that $C$ is built with high accuracy and achieves near-optimal performance when there is no attack present.

To launch a jamming attack, $A$ builds classifier $\hat C$ based on $1000$ observations on the outcome of $T$'s actions by solving a similar optimization as {\it HyperT} in Section~\ref{sec:model} and applies $\hat C$ on the same $500$ test data samples (where $n_{new} = 10$ most recent sensing results are aggregated as one input sample).
We find that this attack significantly jams many transmissions and then decreases the performance of $T$. In particular, the 
normalized throughput is reduced to $22.70\%$ and the success probability is reduced to $23.12\%$.

Moreover, $T$ may update its decision strategy based on feedback (ACK or not) in the retraining process. 
That is, if there is an ACK, $T$ labels the current time slot as idle and if there is no ACK for a transmission, $T$ labels the current time slot as busy.
It is expected that $T$ can improve its classifier $C$ by retraining $C$ with additional data.
However, $A$ can launch an attack by determining when to jam such that some labels are incorrect.
As a result, the updated classifier for $T$ may become worse, i.e., it may make more wrong decisions than $T$'s current classifier. This corresponds to a \emph{causative attack}.

Suppose that classifier $C$ is updated by additional $1000$ training data samples under such an attack.
We find that the performance of the updated classifier drops.
Under this attack, the classifier $C$ is updated as $\tilde C$.
The false alarm and misdetection probabilities are $e_{_{FA}}(\tilde C)=53.10\%$ and $e_{_{MD}}(\tilde C)=31.96\%$, respectively.
The normalized throughput is $46.40\%$ and the success probability is $85.00\%$. Note that the attack on retraining process is confined to the training period, which is typically much shorter than the test period, and therefore such an attack is more energy-efficient and stealthier than attacking the test phase as discussed before.
Table~\ref{table:jamming} summarizes results for jamming attacks in both test and training periods.

\section{Spectrum Poisoning Attack}
\label{sec:sense}
Adversarial deep learning to jam the test phase of spectrum sensing was considered in \cite{Yi2018Milcom}. We first use this attack as a benchmark and then extend this attack to the (re)training phase to improve energy efficiency and stealthiness of the attack. 

\begin{figure}
	\centering
	\includegraphics[width=0.7\columnwidth]{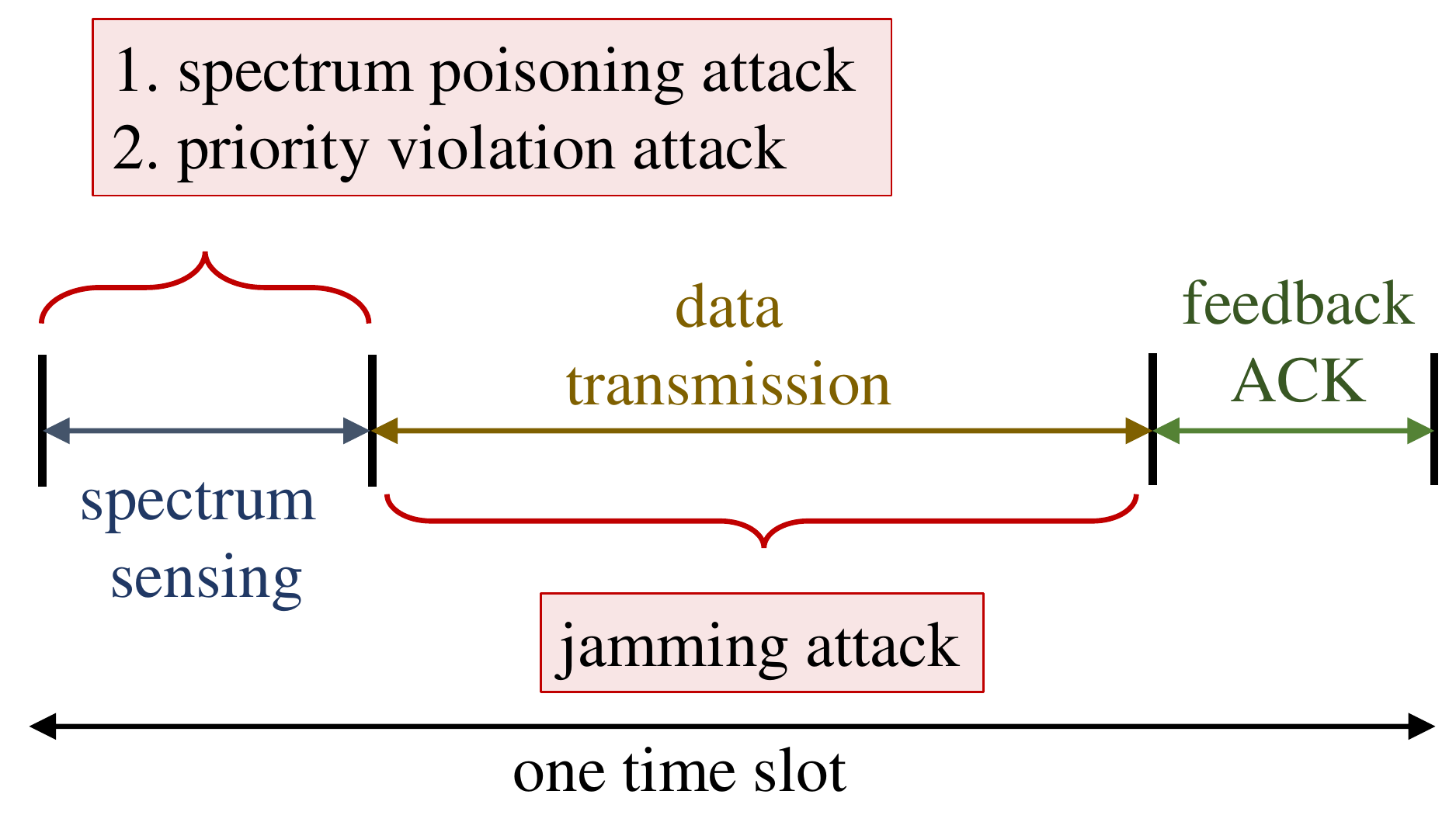}
	\caption{Attacks on different phases.}
\label{fig:frame}
	\vspace{-5mm}
\end{figure}

In the jamming attack, $A$ jams the data transmission phase. 
\rev{In the spectrum poisoning attack, $A$ makes transmissions to change sensing results in the sensing phase (see Figure~\ref{fig:frame}), which are the features used by $T$, and thus mislead the machine learning algorithm of $T$ by providing wrong features.} Hence, $T$ makes a wrong decision, i.e., does not transmit when the channel is idle or transmits when the channel is busy. This corresponds to an \emph{evasion attack}.

We consider the same network setting as in Section~\ref{sec:jam}.
Without an attack, the false alarm and misdetection probabilities of classifier $C$ are measured as $e_{_{FA}}(C)=2.68\%$ and $e_{_{MD}}(C)=1.03\%$, respectively.
Using $C$ to make transmission decisions, the normalized throughput and the success probability are measured as $96.43\%$ and $95.58\%$, respectively. 
Under this attack, the input to $C$ is modified by $A$ and thus the false alarm and misdetection probabilities change to $e_{_{FA}}(C)=91.96\%$ and $e_{_{MD}}(C)=1.29\%$, respectively.
The normalized throughput is reduced to $8.04\%$ and the success probability is reduced to $64.29\%$.  Note  that  the sensing  phase  is  much  shorter  than  the  transmission  phase. Thus, the spectrum poisoning attack consumes much less energy than the jamming  attack, e.g., if   the  length  of  data transmission  is $9$ times  of  the  length  of  spectrum  sensing, spectrum poisoning attack can save energy by $88.89\%$ while the reduction in normalized throughput of $T$ increases by $15\%$ compared to jamming attack. 

$A$ can also provide wrong features in the retraining process.
Under such an attack, the classifier $C$ is incorrectly updated as $\tilde C$.
The false alarm and misdetection probabilities are $e_{_{FA}}(\tilde C)=65.18\%$ and $e_{_{MD}}(\tilde C)=29.64\%$, respectively.
The normalized throughput is reduced to $46.43\%$ and the success probability is reduced to $33.77\%$.
Table~\ref{table:poisoning} summarizes results for spectrum poisoning attack in test and retraining phases.

\begin{table*}
	\caption{\rev{Priority violation attacks based on adversarial machine learning.}}
	\centering
	{\small
		\begin{tabular}{c|c|c|c|c}
			Attack & False alarm & Misdetection & Normalized throughput & Success ratio \\ \hline
			No attack & 19.62\% & 0.83\% & 79.62\% & 98.10\% \\ \hline
			\rev{Priority violation attack} & 88.08\% & 0.42\% & 11.92\% & 96.88\% \\ \hline
			\rev{Priority violation attack on retraining} & 26.54\% & 14.17\% & 74.23\% & 85.78\%
		\end{tabular}
	}
	\label{table:emulation}
	\vspace{-5mm}
\end{table*}


\section{\rev{Priority Violation Attack}}
\label{sec:pu}

\rev{We consider the scenario that the unknown transmitter $B$ is a high-priority, potentially an IoT, user. 
$T$ needs to predict the existence of high-priority user's transmission by machine learning and avoid transmitting at the same time.
We assume that the high-priority user transmits in a time frame with probability $0.2$ and such a transmission takes $5$ time slots.
Once a high-priority transmission is detected, $T$ will back off for $2$ time slots.} We implement this scenario using the same network setting in Section~\ref{sec:jam}.
Due to different background traffic, $T$ needs to train its classifier $C$ and achieves a different performance.
The false alarm and misdetection probabilities of $C$ are $e_{_{FA}}(C)=19.62\%$ and $e_{_{MD}}(C)=0.83\%$, respectively.
The normalized throughput is measured as $79.62\%$ and the success probability is measured as $98.10\%$. \rev{Note that the false alarm probability is larger than that in Section~\ref{sec:jam} or \ref{sec:sense}, which is the cost to protect high-priority transmissions.} This also causes a smaller throughput.

\rev{An adversary $A$ learns the outcome of $T$'s transmissions by deep learning (exploratory attack) and tries to behave like a high-priority user (e.g., replay the high-priority user's signal in the sensing phase) when it predicts a successful transmission of $T$.} This attack is again on the sensing phase (see Figure~\ref{fig:frame}).
\rev{Moreover, due to the backoff mechanism, there is no need to replay the high-priority user's signal in every time slot.
Compared to spectrum poisoning attack, which is an energy efficient attack, we observe that the number of attack actions can be reduced by $46.20\%$, which corresponds to further reduction on energy consumption.} Under this attack, the input to $C$ is changed. Then the false alarm and misdetection probabilities are $e_{_{FA}}(C)=88.08\%$ and $e_{_{MD}}(C)=0.42\%$, respectively.
The normalized throughput is reduced to $11.92\%$. The success probability is $96.88\%$ since only few transmission attempts are made and those are still mostly successful.

When $T$ retrains its classifier, $A$ provides wrong features in the retraining process. 
Under this attack, the false alarm and misdetection probabilities are $e_{_{FA}}(\tilde C)=26.54\%$ and $e_{_{MD}}(\tilde C)=14.17\%$, respectively.
The normalized throughput is $74.23\%$ and the success probability is $85.78\%$.
This attack is not very successful (the normalized throughput is reduced from $79.62\%$ to $74.23\%$ and the success probability is reduced from $99.05\%$ to $85.78\%$) since the number of attack actions is reduced and thus has a limited impact in the retraining process.
Table~\ref{table:emulation} summarizes results for the priority violation attack.

\section{Defense Approach to Adversarial Machine Learning for Wireless Communications}
\label{sec:discuss}

\rev{The results under the three attack scenarios shown in Tables~\ref{table:jamming}, \ref{table:poisoning}, and \ref{table:emulation} demonstrate the success of adversarial machine learning in wireless communications and raise the need of developing effective \emph{defense} approaches to protect machine learning-based IoT systems.
}

Exploratory attack is the basis of subsequent evasion and causative attacks.
The data collected by $A$ for the exploratory attack includes features and labels.
$T$ cannot change features (sensing results) but can change labels by taking different transmit actions.
One approach is randomizing the deep learning network by adding small variations to output labels such that transmit actions can be changed, e.g., transmit in a busy channel or not transmit in an idle channel.
By taking these different (wrong) transmit actions, $T$ aims to fool $A$ and prevents it from building  a good classifier $\hat C$ in the exploratory attack that is used to predict the outcome of $T$'s transmissions.
There is a tradeoff in this defense scheme, i.e., taking many different actions reduces the performance of $T$ although it provides a better protection against the attacks. 
Thus, it is important for $T$ to tune the level of controlled variations added to decisions such that the performance drop of $T$ is limited while the protection against the attacks is still significant.

The above approach, although simple, has a drawback that we cannot selectively make wrong actions to maximize the protection effect.
In this paper, we consider a better approach by noting that deep learning classifier $C$ provides a score to characterize the confidence on classification.
Thus, to maximize the impact of defense actions, wrong actions should be taken when the confidence is high.
In particular, the score $S(s_i)$ is compared with threshold $\tau$ for classification, i.e., $s_i$ is classified as label $0$ if $S(s_i) \le \tau$ and as label $1$ otherwise.
If a sample has a score close to $\tau$, the confidence of classification is low, otherwise the confidence is high.
Based on the classification score in the training data, we first determine two thresholds $\tau_0$ and $\tau_1$ such that $\tau_0 < \tau < \tau_1$ and there are $50\%$ samples with label $0$ and $S(s_i) \le \tau_0$ and $50\%$ samples with label $1$ and $S(s_i) \ge \tau_1$.
Then a sample is regarded to have high classification confidence if $S(s_i) \le \tau_0$ or $S(s_i) \ge \tau_1$.

\begin{figure}
	\centering
	\includegraphics[width=0.8\columnwidth]{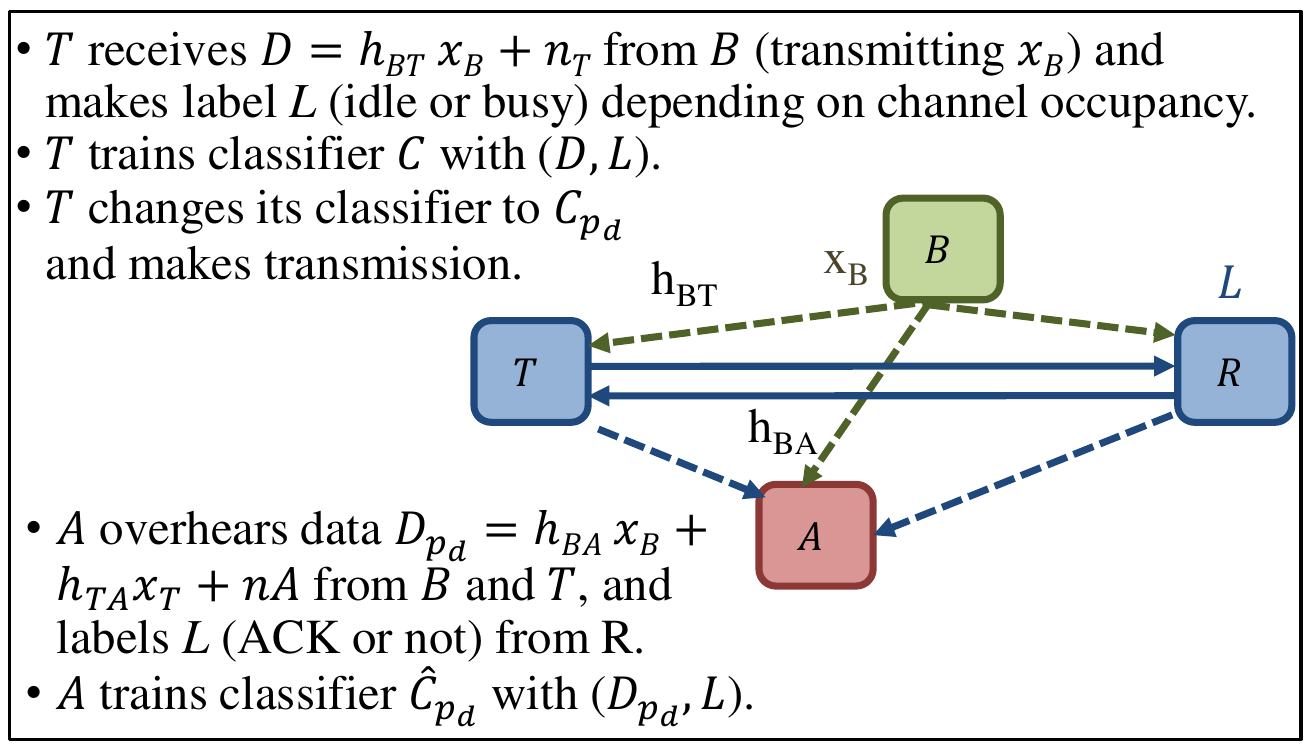}
	\caption{The defense procedure.}
	\label{fig:defense}
\end{figure}

In this defense, we consider a tradeoff on the number of wrong actions and the level of protection against the attacks.
Transmitter $T$ selects the probability of defense actions $p_d$, i.e.,
for each sample with high confidence, $T$ may take a defense action with probability $p_d$.
We can denote the revised classifier of $T$ as $C_{p_d}$.
Adversary $A$ then optimizes its deep learning classifier $\hat C_{p_d}$ to predict transmission outcomes.
$A$ launches an attack using $\hat C_{p_d}$ and under this attack, $T$ applies classifier $C_{p_d}$ and achieves certain performance, e.g., normalized throughput.
By observing this performance, $T$ adapts $p_d$ to improve its performance.
This procedure is shown in Figure~\ref{fig:defense}.
This is a \emph{Stackelberg game}, where $T$ is the leader and $A$ is the follower.
In this game, $T$ aims to maximize the normalized throughput $V^*(p_d)$ by optimizing $p_d$ while $A$ aims to minimize the normalized throughput $V(p_d,H)$ for given $p_d$ by optimizing deep learning parameters $H$, where $V^*(p_d) = \min_H V(p_d,H)$.

This optimization problem can be formulated as 
\begin{eqnarray*}
\mbox{{\it DefenseT}: maximize} && \min_H V(p_d, H) \\
\mbox{subject to} && V(p_d, H) = \mathcal{R}(C_{p_d}, \hat C_{p_d}(H))  \\
\mbox{variable:} && p_d,
\end{eqnarray*}
where $\hat C_{p_d}(H)$ is $A$'s classifier with hyperparameter $H$ when $T$ takes defense actions with probability $p_d$ and $\mathcal{R}(C_{p_d}, \hat C_{p_d}(H))$ is the reward function that measures normalized throughput under classifiers $C_{p_d}$ and $\hat C_{p_d}(H)$.

\begin{figure}
	\centering
	\includegraphics[width=0.8\columnwidth]{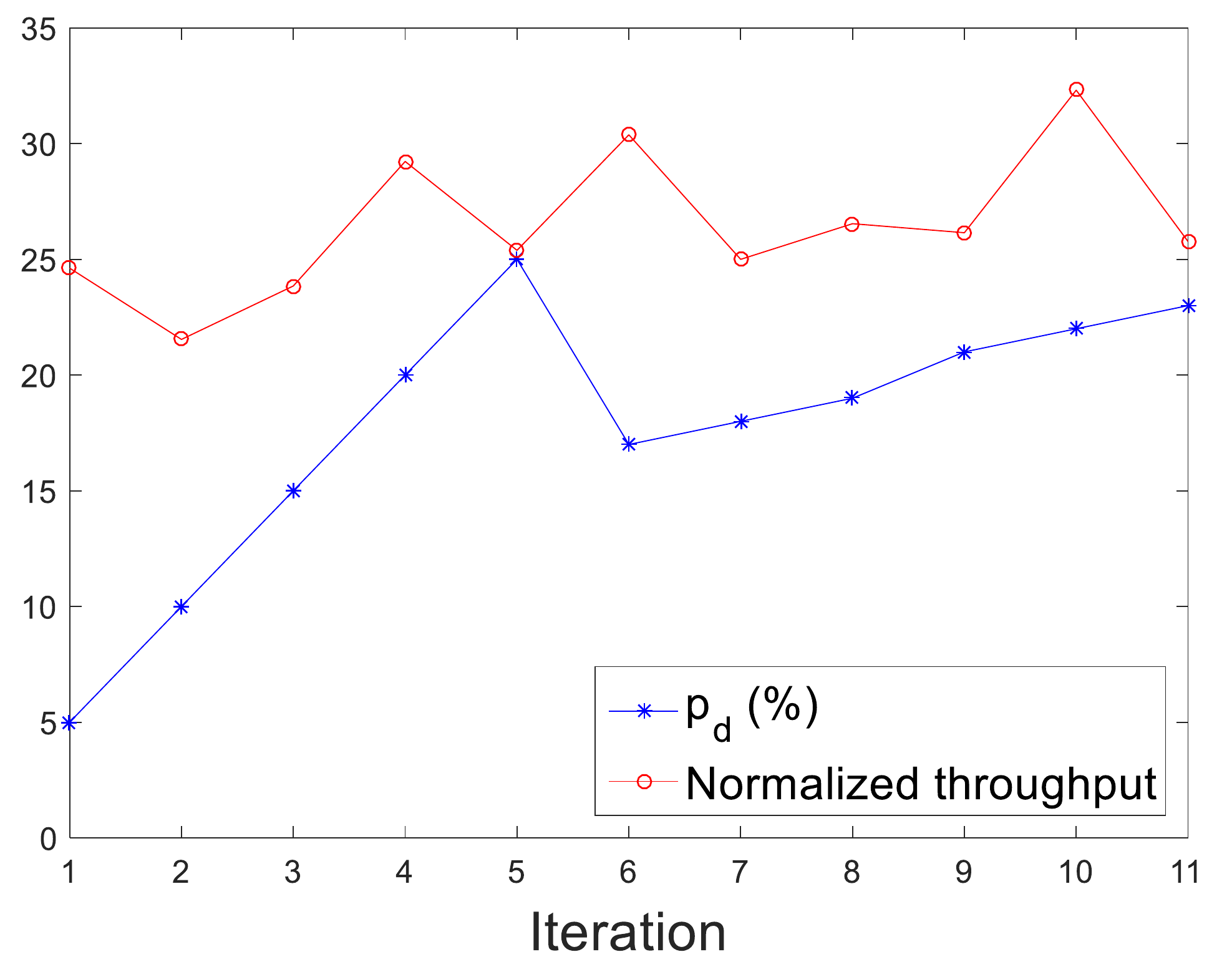}
	\caption{Defense probability $p_d$ and achieved normalized throughput (both measured in $\%$)  during the search process.}
	\label{fig:search}
	\vspace{-5mm}
\end{figure}

{\it DefenseT} targets the exploratory attack and does not depend on the nature of subsequent attacks. 
\rev{In particular, this defense was applied against jamming attacks in
\cite{Yi2018, Tugba2018}. In this paper, we demonstrate the performance of {\it DefenseT} against the priority violation attack and show that it is effective against a broad range of attacks.} Without the defense, the throughput is $11.92\%$ (see Table~\ref{table:emulation}).
Since the optimization process at $A$ is unknown to $T$, $T$ will search for optimal $p_d$ with smaller step size in each round.
In the first round, $T$ takes step size $5\%$ and tries different levels of defense, i.e., $p_d = 5\%, 10\%, \cdots, 25\%$.
The achieved throughputs are $24.62\%, 21.54\%, 23.85\%, 29.23\%$, and $25.39\%$, respectively.
Since $p_d=20\%$ yields the largest throughput, $T$ will further search the region around $40\%$ using a smaller step size $1\%$ in the second round.
That is, $T$ takes $p_d = 17\%, 18\%, 19\%, 21\%, 22\%,$ and $23\%$.
The achieved throughputs are $30.38\%, 25\%, 26.54\%, 26.15\%, 32.31\%,$ and $25.77\%$, respectively.
Since $p_d=22\%$ yields the largest throughput, $T$ will use $p_d=22\%$ as the final solution. The defense increases throughput from $11.92\%$ to $32.31\%$.
Figure~\ref{fig:search} shows $p_d$ value and  normalized throughput over iterations.

\section{Conclusion}
\label{sec:conclusion}

We developed adversarial machine learning techniques to attack and defend IoT systems. We considered an IoT network, where an IoT transmitter senses channel and applies deep learning to predict the channel status (idle or busy) based on the most recent sensing results.
When there are high-priority users, the IoT transmitter with lower priority employs a backoff mechanism to avoid its interference to high-priority user transmissions.
We showed that deep learning can achieve near-optimal throughput.
We then studied adversarial machine learning applied to this setting by applying  exploratory, evasion and causative attacks to jamming, spectrum poisoning and priority violation attacks
The adversary first applies exploratory attack to predict the outcome (ACK or not) of transmitter's decisions.
Then the adversary either changes the test data (evasion attack) or the data in the retraining process (causative attack) to reduce the IoT transmitter's performance. Our results demonstrate new ways to attack IoT systems with adversarial machine learning applied as jamming, spectrum poisoning and priority violation attacks, and report major performance loss in IoT systems. 
Finally, we designed a defense approach based on Stackelberg game and showed its effectiveness in improving the transmitter's performance.

\end{document}